\documentclass[showpacs,preprintnumbers,amsmath,amssymb]{revtex4}

\begin{document}

\title{Second order perturbations in the radius stabilized\\
Randall-Sundrum two branes model. II. \\
\it --- Effect of relaxing strong coupling approximation ---}

\author{Hideaki Kudoh${^{1}}$ }
\email{kudoh@yukawa.kyoto-u.ac.jp}
\author{Takahiro Tanaka${^{2}}$} 
\email{tanaka@yukawa.kyoto-u.ac.jp}

\affiliation{  ${^{1}}$  Department of Physics, Kyoto University, Kyoto 606-8502, Japan 
\\
 ${^{2}}$ Yukawa Institute for Theoretical Physics, 
             Kyoto University, Kyoto 606-8502, Japan}

\begin{abstract}
We discuss  gravitational perturbations in the Randall-Sundrum two 
branes model with radius stabilization. 
Following the idea by Goldberger and Wise for the radius stabilization, 
we introduce a scalar field which has potentials localized on 
the branes in addition to a bulk potential. 
In our previous paper we discussed gravitational perturbations 
induced by static, spherically symmetric and nonrelativistic matter 
distribution on the branes under the condition 
that the values of the scalar field 
on the respective branes cannot fluctuate due to 
its extremely narrow brane potentials.  
We call this case the strong coupling limit.   
Our concern in this paper is to generalize our previous analysis 
relaxing the limitation of taking the strong coupling limit.  
We find that new corrections in metric perturbations 
due to relaxing the strong coupling limit enhance the deviation from the 
4D Einstein gravity only in some exceptional cases.  
In the case that matter fields reside on the negative 
tension brane, the stabilized radion mass becomes very small when 
the new correction becomes large. 
\end{abstract}

\pacs{04.50.+h; 98.80.Cq}
 
\preprint{KUNS-1784, YITP-02-32}

\maketitle

\newcommand{\stac}[2]{ \stackrel{\scriptscriptstyle {#1}}{#2}   }


\section{Introduction}

Understanding the role of extra dimensions has long been a focus of
research. Recent developments in string theory stimulate a
new possibility in a way to realize our Universe, i.e., the ``braneworld,''
The braneworld scenario suggests that our Universe is realized on 
a brane embedded in a higher dimensional spacetime 
\cite{Arkani-Hamed:1998rs,Randall:1999ee,Randall:1999vf}
(see also \cite{Rubakov:1983bb,Visser:1985qm,Akama:1987ig}).

The explicit models introduced by Randall and Sundrum (RS) are 
simple but have attractive features \cite{Randall:1999ee,Randall:1999vf}. 
The two branes model, which was proposed earlier, 
is constructed by orbifold compactification of the 5D anti-de Sitter (AdS) spacetime, in which the two flat branes are on the $Z_2$ symmetric fixed points. The single brane model, which was proposed later, is obtained by pushing the second brane to infinity in the first model.

Since the bulk geometry of these models is warped, the behavior of gravity is not so trivial, and various interesting aspects of gravity in these models have been discussed \cite{Lykken:2000nb,Mukohyama:2001wq,Garriga:2000yh,Tanaka:2000er,Shiromizu:2000wj,Giddings:2000mu,Tanaka:2000zv,Mukohyama:2001ks,Mukohyama:2001jv}.  
One fundamental but remarkable fact is that in the RS single brane model
4D linearized Einstein gravity on the brane is derived from 5D Einstein gravity \cite{Randall:1999vf,Garriga:2000yh,Giddings:2000mu}. 
Also in the RS two branes model,  4D linearized Einstein gravity is restored  \cite{Tanaka:2000er,Mukohyama:2001ks} if 
the distance between branes, which we refer to as radius, is stabilized 
\cite{Goldberger:1999uk,Goldberger:2000un,DeWolfe:2000cp,Goldberger:2000dv,Arkani-Hamed:2001kx,Luty:2001ec,Hofmann:2001cj,Garriga:2001jb,Garriga:2001ar,Flachi:2001bj,Csaki:2000mp,Nojiri:2000bz,Nojiri:2000eb}.  
Although 4D Einstein gravity is approximately recovered
in both models, the corrections arise in different manners. 
In the single brane model, the leading correction to 
Newtonian potential appears in the form of an inverse cubic potential, and therefore the correction is long ranged.  This is because the mass spectrum of the Kaluza-Klein modes is continuous starting with (mass)$^2=0$.  
On the other hand, in the two branes model, the mass spectrum is discrete, and hence the correction becomes short ranged. As was pointed out in Ref. \cite{Tanaka:2000zv}, the leading correction to the metric perturbation can be relatively large in amplitude but it is proportional to the local energy density of the distributed matter on the branes.

The gravity beyond linear perturbations is also an interesting subject to study. For the single brane model,
to study non-perturbative aspects of gravity, many authors have discussed black holes in the braneworld \cite{Gregory:1993vy,Chamblin:2000by,Emparan:2000wa,Dadhich:2000am,Giannakis:2001ss,Chamblin:2001ra,Bruni:2001fd,Horowitz:2001cz,Cadeau:2001tj,Vacaru:2001rf,Vacaru:2001wc,Emparan:2001wn,Kanti:2001cj,Casadio:2001jg,Tanaka:2002rb}.
However, any black hole solutions that can be thought of as a state after gravitational collapse on the brane have not been found yet. On the other hand, there are studies of compact star on the brane \cite{Germani:2001du,Deruelle:2001fb}.
A pioneering work of numerically solving a relativistic star on the brane was done by Wiseman \cite{Wiseman:2001xt}.  
Another direction of research is to study higher order perturbations. 
Second order perturbations have been studied, and 4D Einstein gravity was proven to be restored under certain restrictions \cite{Giannakis:2001zx,Kudoh:2001wb}.

For the two branes model, the recovery of 4D Einstein gravity is also concluded in our previous paper 
\cite{Kudoh:2001kz} under some assumptions.
However, the mechanism for the recovery in each model is not so clearly understood as in the case of linear perturbations.  

The basic assumptions taken in the second order perturbations of the two branes model are that the radius is stabilized by the mechanism proposed by Goldberger and Wise (GW) \cite{Goldberger:1999uk}, and that the matter distribution on the brane is static, spherically symmetric and non-relativistic. 
In the GW model for radius stabilization, a bulk scalar field is introduced. This scalar field has potentials localized on the branes as well as a bulk potential. 
A further technical assumption taken in the previous analyses \cite{Tanaka:2000er,Kudoh:2001kz} is that the values of the scalar field 
on the respective branes are stuck to fixed values due to extremely narrow brane potentials.  
We refer to this simplified setup as the strong coupling limit.   
In this paper we generalize our previous analyses of linear and second order perturbations \cite{Kudoh:2001kz}  
relaxing the restriction of the strong coupling limit 
\footnote{Recently, Mukohyama gave a similar analysis of linear perturbations by a completely different approach \cite{Mukohyama:2001jv}.},
and study whether there arise observable effects and/or no pathological feature in the metric perturbations.

The paper is organized as follows. In Sec.~\ref{sec:review} we briefly review the formulation developed in Ref.~\cite{Kudoh:2001kz}, which we call paper I, summarizing the notations and the basic equations.  
We do not give the explicit form of all the necessary formulas that are already presented in paper I to avoid repetitions of rather lengthy expressions. 
We quote the equations in paper I such as (I-1.1).  
Throughout this paper, we use the same notations as those in paper I except for the subscripts ${`` \mathrm{pse}" }$ and ${``\lambda,"}$ which are introduced in Eq. (\ref{eq:Y=Y_0+Y_S}).  
In Sec.~\ref{sec:finite coupling}, we study the corrections caused by the finiteness of the coupling strength.  
Section \ref{sec:summary} is devoted to the summary.

\section{Brief review of the formulation}
\label{sec:review}

In this section, we briefly review the formalism and the results presented in paper I.  
We consider second order perturbations in the RS two branes model assuming that matter distribution is confined on one of the branes and it is static and spherically symmetric.

\subsection{Notation and assumption}

The Lagrangian for the scalar field $\tilde\varphi$ introduced for the radius stabilization is  
\begin{equation}
 {\cal L}= - \frac{1}{2} \tilde  g^{ab} 
     \tilde\varphi_{,a} \tilde\varphi_{,b}
          -V_B(\tilde\varphi)
          -\sum_{\sigma=\pm} V_{(\sigma)}(\tilde\varphi)
              \delta(y-y_{\sigma}) \,,
\label{eq:Lagrangian}
\end{equation}
where $V_B$ and $V_{(\pm)}$ are the bulk potential and the potential on each brane, respectively.

In analyzing metric perturbations in the bulk, we use the ``Newton gauge,'' in which   
\begin{eqnarray}
 ds^2 
     = e^{2 Y } dy^2 + a^2 
\left[- e^{ A - \psi } dt^2
      + e^{ B - \psi } dr^2
      + e^{ C - \psi } r^2 d\Omega^2 
\right], 
\label{eq:metric}
\end{eqnarray}
where the metric functions $A, B, C, Y$, and $\psi$ depend only on $y$ and $r$, and $a(y)$ is the warp factor that is determined by solving the background equations (I-2.10).
The metric functions and the scalar field are expanded up to the second order as 
\begin{eqnarray}
  A(r,y) &=& A^{(1)}(r,y) +A^{(2)} (r,y) ,
\\
 \tilde\varphi (r, y) 
 &=& \phi_0(y) + \varphi^{(1)}(r,y)+\varphi^{(2)}(r,y) ,
\end{eqnarray}
where $\phi_0$ represents the background scalar field configuration, which depends only on $y$. The metric functions $A$, $B$, and $C$ are related to each other by the relations (I-2.5) so that they compose the transverse-traceless part at the linear level.

Our interest is the gravity induced by non-relativistic matter fields confined on the positive and the negative tension branes, which are located at $y=y_+$ and $y=y_- (> y_+)$, respectively. 
The energy-momentum tensors of these fields are assumed to be given in the perfect fluid form as
\begin{eqnarray}
    T_{{\pm} \nu }{}^{\!\mu} = 
    a_\pm^{-4} \,
    {\mathrm{diag}} 
    \{- \rho_{\pm}, P_{\pm}, P_{\pm}, P_{\pm} \}.
\label{eq:EM tensor}
\end{eqnarray}
The warp factor $a_\pm:=a(y_\pm)$ in the definition of the energy-momentum tensors is incorporated so that $\rho$ and $P$ become the physical quantities measured by using the metric on the positive tension brane.

The 5D Einstein equations give four independent perturbation equations.
The two equations are the constraint equations for $\psi^{(J)}$ and $\varphi^{(J)}$, which relate them to $Y^{(J)}$ [Eqs.~(I-2.12) and (I-2.13)]:
\begin{eqnarray}
   && \psi^{(J)} (r,y) = Y^{(J)} + \epsilon^{(J)} \Delta^{-1} S_\psi  \,,
\label{eq: phi^J=Y^J+...}
\\
   &&   \varphi^{(J)} (r,y)
    = \frac{3}{2\kappa \dot \phi_0 a^2}  \partial_y( a^2 Y^{(J)})
    + \frac{3}{2\kappa \dot \phi_0} \epsilon^{(J)} \Bigl[
      S_\varphi 
    + \partial_y \Delta^{-1} 
  S_\psi 
    \Bigr] \,,
    \label{eq:d varphi = Y^J +...}
\end{eqnarray}
where we have introduced $\epsilon^{(J)}$ defined by
$\epsilon^{(1)}=0$ and $\epsilon^{(2)}=1$.
Hence, as for the scalar type perturbation, once we solve the perturbation $Y^{(J)}$, other variables $\psi^{(J)}$ and $\varphi^{(J)}$ are also known. 
The other two equations are the master equations for $A^{(J)}$ and
$Y^{(J)}$ [Eqs.~(I-2.14) and (I-2.15)]:
\begin{eqnarray}
   &&  \left[   {1\over a^2}\partial_y a^4\partial_y {1\over a^2} +\frac{1}{a^2} \Delta 
       \right] (a^2A^{(J)})= \epsilon^{(J)} S_A \,,  
\label{eq:Ein-A^J}   
\\
   && \left[
     a^2\dot\phi_0^2\partial_y{1\over a^{2}\dot\phi_0^2}
      \partial_y  
    - {2\kappa\over 3} \dot\phi_0^2  + {1\over a^2} \Delta 
      \right] (a^2Y^{(J)} ) = \epsilon^{(J)} S_Y\,.
\label{eq:Ein-Y^J}
\end{eqnarray}

To solve the master equations, we must specify the boundary conditions on the branes.
The boundary condition for $A^{(J)}$ is given by Israel's junction condition, whereas the boundary condition for the scalar type perturbation is derived by integrating the equation of motion for the scalar field across the branes. It is well known that these junction conditions are easily obtained in Gaussian normal coordinates $\bar x^a$ in which a brane is located at $\bar y=\mathrm{const}$ hypersurface. 
Here we associate an over-bar with quantities written in Gaussian normal coordinates.  The boundary conditions in the Newton gauge are derived by applying infinitesimal gauge transformations $\bar x^a = x^a + \xi^a$ to those written in Gaussian normal coordinates.
These transformations are described in Sec.~II B in paper I.

The boundary condition for the scalar type perturbation is given by [Eq.~(I-2.37)] 
\begin{eqnarray}
 \varphi^{(J)} - \dot\phi_{0 }   
 \stac{(J)~}{\hat\xi_{\pm}^y}
 = 
 \frac{\lambda_{\pm}}{2}
 \left(
  \mp \frac{3}{ \kappa a_\pm^2 \dot\phi_{0 }}
     \Delta  Y^{(J)}
   + \epsilon^{(J)}  S^{ \pm }_{jun} \right) 
  \quad
  ({\mathrm{at~}} y=y_\pm)
   \,,
\label{eq:Jun of varphi}
\end{eqnarray}
where we have introduced a coupling constant 
\begin{equation}
   \lambda_{\pm}
   :=\frac{2}{ V''_{(\pm)}\mp 2(\ddot\phi_{0\pm}/\dot\phi_{0\pm})} \,. 
\label{eq:lambdadef}
\end{equation}
An over-dot denotes differentiation with respect to $y$, and
$\phi_{0\pm}:=\phi_{0}(y_\pm)$. 
Note that the explicit expression for 
$S_{jun}$ is given in Eq.~(I-2.39). 
The parameter of the gauge transformation in $y$ direction, ${\hat\xi^y_{\pm}}(r)$, is a function of $r$, and it is related to the energy-momentum tensor on the corresponding brane by Eq.~(I-2.34).

In paper I, the problem was simplified by taking the strong coupling limit. 
The condition of the strong coupling is that 
$V_{(\pm)}''$ is sufficiently large. 
Taking this limit, we obtain 
\begin{eqnarray}
    \lambda_{\pm} \to 0 .
\label{eq:large limit}
\end{eqnarray}
In this paper we discuss the effects due to the terms that arise by considering non-vanishing $\lambda_\pm$.  
For brevity, we refer to these terms as interaction terms.

\subsection{Derivative expansion}

Formal solutions for perturbation equations (\ref{eq:Ein-A^J}) and (\ref{eq:Ein-Y^J}) were derived in
paper I. 
The transverse-traceless perturbations $A$ are decomposed to massless zero mode $A_0$ and massive mode $A_S$.
Using the zero mode truncation approximation, the long-ranged part of the transverse-traceless perturbation $A_0$ is evaluated [Eq.~(I-3.5)]. The remaining part $A_S$ [Eq.~(I-3.9) and (I-3.12)], which arises due to the contribution from massive Kaluza-Klein modes, is evaluated by using a derivative expansion method.  In this method, we expand perturbation variables in terms of the expansion parameter $(H r_\star)^{-1} \sim H^{-1} \partial_r$ assuming that the typical length scale $r_\star$ of perturbations is much longer than the 5D curvature scale $H^{-1}:=(\dot a/a)^{-1}$. 
It is important to stress that this derivative expansion method is valid only when the mass of the first excited mode is sufficiently large. In the limit $y_- \to \infty$, the excited mass spectrum becomes continuous, and therefore the derivative expansion method is no longer valid. 
(See Secs.~III B.1 and IV B in paper I for more details.)

As for the scalar type perturbation, there is no zero mode owing to the stabilization mechanism \cite{Tanaka:2000er}. 
To discuss the contributions from massive modes, we expand the perturbation variables by using the derivative expansion.  Although the massive modes seem to give only the short-ranged part, it turns out that the formal solution of the lowest order in the expansion includes long-ranged metric perturbations. Hence, we refer to this part as the pseudo long-ranged part. The formal solution of the next order, which is obtained by an iteration, is referred to as the short-ranged part $Y_S$.  
(See Sec.~III C in paper I.)

The pseudo-long-ranged part includes the contribution from the interaction terms that are higher order in derivative expansion as the short-ranged part. Hence we further divide the pseudo-long-ranged part into two pieces; the contribution from the interaction terms, which we denote $Y_{\lambda}$, and the remaining terms, which we denote $Y_{ \mathrm{pse}}$.
We quote the explicit expression for the pseudo-long-ranged part [see Eqs.~(I-3.21) and (I-B.6)], 
\begin{eqnarray}
 Y^{(J)} &=& Y^{(J)}_{ {\mathrm{pse}} } + Y^{(J)}_{\lambda} + Y^{(J)}_{S} , 
\label{eq:Y=Y_0+Y_S}
\cr
 \Delta Y^{(J)}_{\mathrm{pse} } (r, y_{\pm})
   &=& 
  - \frac{\kappa N}{3} \sum_{\sigma=\pm} a_{\sigma}^4 
    T^{(J)}_{\sigma}  
    \mp \frac{\kappa H}{3} a_{\pm}^2 T_{\pm}^{(J)}
 - \epsilon ^{(J)} 
 \Bigl(  S_\psi + 2H a_{\pm}^2 S_{\xi}^{\pm}
 \Bigr) 
\nonumber
\\
&&
   - 2N \epsilon^{(J)} \biggl[
     \sum_{\sigma=\pm} \sigma a_{\sigma}^4
     S_{\xi}^{\sigma} 
   + \int^{y_{-}}_{y_{+}} dy 
 \Bigl(
      a^2 v_{\pm} \Delta {\mathbb S}_\varphi 
    + \frac{3u_{\pm}} {2\kappa \dot \phi_0^2} \Delta {\mathbb S}_Y 
    -a^2 S_\psi 
  \Bigr)
\biggr] \,,
\label{eq:Y^J_[0] y_+-}
\cr
\Delta  Y^{(J)}_{ \lambda } (r, y_\pm)
&=&
 - 2N  \Delta  \left(  
    \sum_{\sigma= \pm} \sigma  \frac{ L_\sigma^{(J)} }{H_\sigma}
  + \frac{ L_\pm^{(J)} }{a^2_\pm N} 
  \right) \,,
\label{eq:Y^J_[0lambda] y_+-}
\end{eqnarray}
where  
\begin{eqnarray}
 L_\sigma^{(J)}(r) = H(y_\sigma)
   \biggl[ 
    - \sigma \frac{3 \lambda_{\sigma}}{2\kappa\dot\phi_0^2} \Delta  Y^{(J)}
   + \epsilon^{(J)} a_{\sigma}^2
  \Bigl( 
    \frac{ \lambda_{\sigma}}{2\dot\phi_0} S^{\sigma}_{jun} 
  - \frac{ \ddot\phi_0  \varphi^2}{2 \dot\phi_0^3} 
  + \frac{3 \varphi  \Delta  Y }
      {2\kappa a^2 \dot\phi_0^3}
  \Bigr) 
   \biggr] _{y=y_{\sigma}}\,.
\label{eq:L_sigma}
\end{eqnarray}
Here $N$ is the normalization factor defined by 
\begin{eqnarray}
N^{-1} := 2\int_{y_{+}}^{y_{-}} a^2 dy,  
\label{eq:N}
\end{eqnarray}
and $u_\pm$ is given by 
\begin{equation}
 u_{\pm}(y) :=1-2 H v_{\pm}, 
\quad
    v_{\pm}(y) :=\frac{1}{a ^2}\int_{y_{\mp}}^y a^2 dy' \,.
    \label{eq:def. v}
\end{equation}
We mention that the source term $S^{\pm}_{jun}$ that is defined in Eq.~(\ref{eq:Jun of varphi}) contains $V''_{(\pm)}$ and $V'''_{(\pm)}$, and hence $\lambda_\pm S^\pm_{jun}$ does not vanish even in the limit (\ref{eq:large limit}). However, it was shown that the combination that appears in Eq.~(\ref{eq:L_sigma}) vanishes in the strong coupling limit. Namely, we have 
\begin{eqnarray}
    L^{(J)}_{ \pm} \approx 0 \quad (\mathrm{for~} \lambda_\pm \to 0 )\,. 
\label{eq:L=0}
\end{eqnarray}

Note also that, in Eq.~(\ref{eq:Y^J_[0lambda] y_+-}), the number of $r$ derivatives on the right hand side is larger by two than that on the left hand side. 
This fact is manifest for linear perturbations, and it also turns out to be the case for the second order. Hence, when we evaluate $Y_{\lambda}^{(J)}$ iteratively, the leading term with respect to gradient expansion is obtained by substituting pseudo-long-ranged part 
$Y_{\mathrm{pse}}^{(J)}$ into  $Y^{(J)}$ on the right hand side of Eq.~(\ref{eq:L_sigma}).  Although the short-ranged part $Y_S^{(J)}$ also contains interaction terms, we do not discuss them in this paper because they are even higher order in $r_{\star}^{-2}$.

At the linear level, the transverse traceless part composed of 
$A$, $B$, and $C$ does not have interaction terms. 
The parameter of the gauge transformation $\stac{(1)}{\hat\xi^y_\pm}$ is also unaltered by the effect of non-vanishing $\lambda_\pm$.  
On the other hand, $\varphi^{(1)}$ is related to $Y^{(1)}$ and changes according to the change in $Y^{(1)}$.

By repeating the derivation of the expression for the temporal component of the metric perturbation induced on the branes (I-5.1) starting with the equation for gauge transformation (I-2.25), we will find Eq.~(I-5.1) is slightly modified as 
\begin{eqnarray}
  \Delta \bar A^{(J)} _{0 }(r,y_{\pm})
   &=&  8 \pi G 
      (\rho ^{(J)}_{\pm} + 3P ^{(J)}_{\pm}  )
 - {\Delta} 
   \Bigl[
     {\hat\xi_{ \pm }^y} \bar{A} _{\pm,y} 
   + {\hat\xi_{ \pm }^r} \bar{A} _{\pm,r}
   + \dot H ( {\hat\xi_{ \pm }^y})^2 
    \Bigr]
   + 2N \sum_\sigma \sigma a_{\sigma}^4 (S_{\xi}^{\sigma} - S_{\Sigma}^{\sigma})
\nonumber
\\
& &  
 +  {2N}  \int^{y_{-}}_{y_{+}} dy 
  \biggl[ 
    a^2 ( S_{A\pm} - S_{\psi\pm})
  + a^2 v_{\pm} \Delta {\mathbb S}_{\varphi\pm} 
  + \frac{3u_{\pm}}{2 \kappa \dot \phi_0^2}  
    \Delta {\mathbb S}_{Y\pm}
\biggr] 
-
  \Delta Y^{(J)}_{ \lambda }
\,.  
\label{eq:bar A_0}
\end{eqnarray}
For the spatial components, it is convenient to take the isotropic gauge.  We simply quote Eq.~(I-3.15) in which the isotropic gauge is taken in $J$th order:
\begin{eqnarray} 
   \Delta \bar B^{(J)}  (r,y_\pm )
   = - \frac{1}{2} \Delta \bar A^{(J)}  
 \mp \frac{\kappa H}{2}a^2_\pm T^{(J)}_\pm
 - \frac{3}{2} \Delta Y^{(J)}
 - \frac{3}{2} \epsilon^{(J)}
\left(
    \frac{2}{3}\Delta  S_B 
  + S_\psi + 2Ha^2_\pm S^\pm_\xi
\right) \,.
\label{eq:bar B^2}
\end{eqnarray}
The explicit additional term in these equations is only the last term
in Eq.~(\ref{eq:bar A_0}), but there are implicit changes through $Y^{(J)}$, $\varphi^{(J)}$, and  
$\bar A^{(J)}$. 
The explicit expressions for the source terms 
$S_*$ and ${\mathbb S}_*$ are given in 
Eqs.~(I-2.16), (I-2.18), (I-2.19), (I-2.21), (I-2.33), (I-2.35) and  (I-3.14).

\subsection{Corrections}

As we have done in paper I, we assume that matter fields reside on one of the two branes.  By this simplification, the sum of the Newton potentials $\Phi_\pm $, which are defined by $\Delta \Phi_\pm (r) = \kappa N \rho_\pm^{(1)}/2$, is replaced as 
\begin{eqnarray}
   \sum_\sigma \Phi_\sigma 
   \to 
   \Phi_\pm  
   \quad   (y=y_{\pm}) \,.
\label{eq:simp} 
\end{eqnarray}
Then, the long-ranged part of the transverse traceless metric perturbations is given by (I-4.7) and (I-4.8) as
\begin{eqnarray}
  [\bar A^{(1)}_{0\pm}(r, y_\pm)]_{SC} =  
  - [\bar B^{(1)}_{0\pm} (r, y_\pm)]_{SC} =  2\Phi_{\pm} \,, 
\end{eqnarray} 
where $[\cdots]_{SC}$ means the quantity in the $\lambda_\pm\to 0$ limit.
The index $\pm$ associated with the metric functions (not with $\lambda$ and $L$) specifies the side on which matter fields are distributed.

To obtain an approximate estimation for the short-ranged part, we assume that the back reaction of the bulk scalar field to the background geometry is weak; 
${|\dot H|}/{H^2}  \ll 1$.
In this case the metric approximately takes the AdS form 
\begin{eqnarray}
   a (y) \approx e^{-|y|/\ell}  \,,
\label{eq: a=exp}
\end{eqnarray}
where $\ell$ is the curvature radius of the bulk. 
For later convenience, we evaluate $N$ in this weak back reaction case. It is approximately given by 
\begin{eqnarray}
   N^{-1} \approx -  \frac{1}{H_+} 
  \left( 1 + \frac{\dot H_+}{2H_+^2} \right) \,,
\label{eq:Ninv}
\end{eqnarray}
taking into account the fact that the integral (\ref{eq:N}) is dominated around $y=y_+$. The second term in the round brackets is the leading order correction due to the back reaction. 
Hereafter we set $a_{+}=1$.

Under the assumption of weak back reaction, it was shown that 
\cite{Tanaka:2000er,Kudoh:2001kz}
\begin{eqnarray}
 \bigl[  A_{S\pm}^{(1)}(r,y_\pm)  \bigr]_{SC} =
\beta_\pm   O(r_\star^2 \Delta \Phi_\pm) \,,
\label{eq:A - KK corr}
\end{eqnarray}
where the suppression factors are 
\begin{eqnarray}
   \beta_+ = \frac{\ell^2}{r_\star^2}\,, \quad\quad
   \beta_- = \frac{\ell^2}{a^4_-  r_\star^2} 
    =\Bigl( \frac{ 0.1 {\mathrm{mm}} }{r_\star} \Bigr)^2
    \Bigl( \frac{ 10^{-16} }{a_-}  \Bigr)^4
    \Bigl( \frac{ \ell}{\ell_{Pl}}  \Bigr)^2\,.
\label{eq:suppres factor 1}
\end{eqnarray}
The short-ranged part for the scalar type perturbation is similarly suppressed as 
\begin{equation}
 \bigl[  Y_{S\pm}^{(1)}(r,y_\pm)  \bigr]_{SC} =
 \frac{a_-^2\beta_\pm}{m_S^2\ell^2} 
 O(r_\star^2 \Delta \Phi_\pm) \,,  
\end{equation}
where $m_S^2$ is the mass squared of the so-called radion, which is the mode corresponding to the radius fluctuation, in the strong coupling limit. 
To be precise, the radion is defined by the mode with the lowest mass eigenvalue in the scalar type perturbation. 
Since $m_S^2 \lesssim  O(a_-^2 \ell^{-2})$, which is given by 
Eq.~(\ref{eq:mass}) below with $\lambda_\pm=0$, the corrections from the scalar type perturbation tend to be larger than those from the Kaluza-Klein modes in the tensor type perturbation.

On the positive tension brane, the short-ranged part is suppressed when the typical length scale $r_\star$ of perturbations is much larger than
$a_{-} m_S^{-1} $. 
On the negative tension brane, the short-ranged part is suppressed for $r_\star \gtrsim  0.1 (a_-/m_S\ell)$ mm, where the ratio $a_+/a_-$ is set to $10^{16}$, the value to solve the hierarchy between Planck and electroweak scales. 
One may think that this effect is potentially observable especially on the negative tension brane. 
As we have mentioned in the introduction, however, the short-ranged part is proportional to the local matter energy density $\rho_\pm^{(1)} \propto \Delta \Phi_\pm$. Therefore, the short-ranged part dose not contribute to the force outside the matter distribution, and it is not observed as a change of the Newton's law. 
Not the force but the change of the metric perturbation due to the short-ranged part becomes significant on the negative tension brane only when $\rho^{(1)}_- \gtrsim (m_S^2\ell^2/a_-^2) O (\textrm{TeV}^4)$.

Extending the linear analysis mentioned above, second order perturbations were investigated in paper I.  In the case that matter fields are confined on the negative tension brane, the long-ranged part at the second order contains corrections to 4D Einstein gravity with relative amplitude of $O(\beta_-/a_-^2)$ compared to the ordinary post-Newtonian terms.
The relative amplitude of the corrections at the second order looks less suppressed by an extra factor of $1/a_-^{2}$ than that at the linear order.
However, in the strong coupling limit, it was shown that these enhanced corrections are completely canceled by the contributions from the short-ranged part, and the recovery of 4D Einstein gravity was confirmed. 
In the case that matter fields are confined on the positive tension brane, the corrections in second order perturbations are relatively $O(\beta_+/a_-^{2})$ compared to the usual post-Newtonian terms.  
However, the appearance of the enhancement by the factor of $1/a_-^2$ is very likely to be an artifact due to the gradient expansion method. Since the condition that the typical length scale of spatial gradient is larger than that of the change in the fifth direction becomes $(\ell^2/a^2_- r_\star^2) = (\beta_+/a^2_-)\ll 1$ near the negative tension brane, $\beta_+/a^2_-$ appears as an expansion parameter.

\section{Contributions due to non-vanishing $\lambda_\pm$}
\label{sec:finite coupling}

In the analysis taking the strong coupling limit, we neglected the terms that vanish when $\lambda_\pm$ is set to zero. 
Here we consider the effect of non-vanishing $\lambda_\pm$ to remove this technical limitation. 
We first discuss linear perturbations, and after that we study second order perturbations.

\subsection{Mass spectrum}

Before discussing metric perturbations induced by matter fields, we study the change of mass eigenvalues for the scalar type perturbations due to the effect of $\lambda_\pm$. 
Setting $\hat\xi_\pm^y =0$ in 
Eq.~(I-2.41) with $J=1$, we obtain an eigenvalue equation for 
$Y^{(1)}$ as 
\begin{equation}
\left[a^2\dot\phi^2_0\partial_y \frac{1}{a^2\dot\phi_0^2}\partial_y
  a^2 -\frac{2\kappa}{3} a^2\dot\phi_0^2 
 +m^2\left(1+2\sum_\sigma \lambda_\sigma \delta (y-y_\sigma)\right)\right] 
  Y^{(1)}=0, 
\label{eq:eigenY}
\end{equation}
where the 4D Laplacian operator $\Delta$ was replaced with the mass squared $m^2$. 
A general solution for small $m^2$ ignoring boundary conditions 
at $y=y_{\pm}$ was approximately constructed in Ref. \cite{Tanaka:2000er}  as 
\begin{eqnarray}
  Y^{(1)} & = & \frac{{\cal N}}{a^2}+\cdots\,, 
\cr
    \frac{2\kappa \varphi^{(1)}}{3\dot\phi_0} 
 &=&
   \frac{1}{a^2\dot\phi_0^2}\partial_y (a^2 Y^{(1)})
     =
{\cal N}\left[\int_{y_+}^{y} \left(
             \frac{2\kappa}{3 a^2 }
            -\frac{m^2}{a^4 \dot\phi_0^2 }\right)dy'
          +c \right]+\cdots\, , 
\label{eq:expansion}
\end{eqnarray}
where $c$ is a constant and ${\cal N}$ is a normalization constant. 
Imposing the boundary condition (\ref{eq:Jun of varphi}) on this approximate solution, we find that $c=-\lambda_+m^2/\dot\phi_{0+}^2a^4_+$ and a mass eigenvalue close to zero is given by 
\begin{eqnarray}
  m^2 \approx \left(\frac{2\kappa}{3}\int_{y_+}^{y_-}
            \frac{dy'}{a^2 }\right)
      ~\left(\int_{y_+}^{y_-}
            \frac{dy'}{a^4 \dot\phi_0^2 }+\sum_\sigma \frac{\lambda_\sigma}
             {a^4_\sigma\dot\phi_{0\sigma}^2}
   \right)^{-1} \,.
\label{eq:mass}
\end{eqnarray}
Using Eq.~(I-B22) with the assumption of weak back reaction, the above expression is approximately rewritten as 
\begin{eqnarray}
 m^2 \approx m_S^2
           \left[1 
     + \frac{3}{2} \sum_\sigma \alpha_\sigma \left(
                 \frac{a_-}{a_\sigma}\right)^4
                \left(\frac{m_S^2}
                {a_-^2 N^2} \right) \right]^{-1}\, ,
\label{eq:m2small}
\end{eqnarray} 
where we have introduced non-dimensional parameters related to the coupling
$\lambda_\pm$ by 
\begin{eqnarray}
 \alpha_\pm :=  \frac{2N^3 \lambda_\pm}{ \kappa \dot\phi_{0\pm}^2 } 
             = - \frac{2N^3 \lambda_\pm}{ 3 \dot H_\pm } \,. 
\label{eq:def. alpha}
\end{eqnarray}
When $\lambda_+ \agt m_S^{-2}\ell^{-1}$, the expansion
(\ref{eq:expansion}) is no longer valid since the correction that comes from the constant $c$ becomes larger than ${\mathcal N}/a^2$.
In such cases, instead, we can consider the large 
$\lambda_+$ limit keeping $m^2\lambda_+$ finite. 
Since we can neglect $m^2$ term in the bulk when $m^2 \ll m_S^2$, we find that an approximate solution to the above equation is given by $Y^{(1)} \approx u_+(y)$ in this limit. From the condition (\ref{eq:Jun of varphi}) at the boundary 
$y=y_+$ the 
mass eigenvalue corresponding to this mode is determined as 
\begin{equation}
 m^2=\frac{2\dot H_+ v_+(y_+)}{\lambda_+ u_+(y_+)}
   \approx \frac{2}{\lambda_+ \ell},  
\label{eq:m2limit}
\end{equation}
where we used Eq.~(I-B3) with the aid of Eq.~(\ref{eq:Ninv}).

When $\lambda_+$ (or $\lambda_-$) takes a large negative value, we can see that the above mode of small mass becomes tachyonic, and hence such a model is manifestly unstable. 
As we decrease $|\lambda_\pm|$ starting with $ \lambda_\pm =-\infty$, the absolute value of $m^2$ in the expression (\ref{eq:m2small}) increases from $m^2=0$ and diverges to $|-\infty|$ when $\lambda_\pm$ is a certain negative value $\Lambda_\pm$, which depends on the details of stabilization model, e.g., $\dot \phi_{0\pm}^2$. 
Then the mass eigenvalue returns from $+\infty$ to $m_S^2$ as 
$\lambda_\pm$ increases from $\Lambda_\pm$. 
However, this does not directly indicate that the tachyonic mode disappears for $\lambda_\pm$ $> \Lambda_\pm$.
This is because Eq.~(\ref{eq:m2small}) is no longer valid for relatively large $|m^2|$. 
For large $\nu^2:=-m^2$, one can solve Eq.~(\ref{eq:eigenY}) by using the WKB approximation as 
\begin{equation}
 Y^{(1)}\approx C_1 e^{\nu \int^y dy'/a(y')} +
          C_2 e^{-\nu \int^y dy'/a(y')}\,, 
\end{equation}
where $C_1$ and $C_2$ are constants. 
Imposing the boundary conditions (\ref{eq:Jun of varphi}) at $y=y_\pm$, we obtain 
$(\lambda_+\nu-a_+)(\lambda_-\nu - a_-) 
 -\exp[2\nu \int_{y_+}^{y_-}a^{-1} dy]  
  (\lambda_+\nu + a_+) (\lambda_- \nu + a_-) \approx 0$. 
Without loss of generality we can assume that $\nu$ is positive. 
Then the exponential factor $\exp[2\nu \int_{y_+}^{y_-}a^{-1} dy ]$ is very large, and hence a solution to the above equation is approximately obtained when either $\lambda_+$ or $\lambda_-$ is negative as 
\begin{equation}
\nu \approx \max (- a_+\lambda_+^{-1}, - a_- \lambda_-^{-1}). 
\end{equation}
This eigenmode is the anticipated tachyonic mode, which remains to exist for any small negative value of $\lambda_+$ or $\lambda_-$. 
Although the values of $\lambda_\pm$ depend on the details of the stabilization model, it is natural to consider the case in which  $|\lambda_\pm|$ is less than or equal to $O(\ell)$, and $\lambda_\pm$ must be positive.  To conclude, we find that the model has tachyonic mode if $\lambda_+$ or $\lambda_-$ is negative.

\subsection{Linear perturbation}

From Eq.~(\ref{eq:bar A_0}), the temporal component of the linear perturbation is given by 
\begin{eqnarray}
  \bar A^{(1)} _{0\pm}(r,y_{\pm})
   =  \bigl[ \bar A^{(1)} _{0\pm}\bigr]_{SC} 
     -  Y^{(1)}_{\lambda}.  
\label{eq:bar A_0^1}
\end{eqnarray}
As for the spatial part, we obtain from Eqs. (\ref{eq:bar B^2}) and (\ref{eq:bar A_0^1}) 
\begin{eqnarray}
  \bar B^{(1)} _{0\pm}(r,y_{\pm})
   =  \bigl[ \bar B^{(1)} _{0\pm}\bigr]_{SC}
     -  Y^{(1)}_{\lambda}.  
\label{eq:bar B_0^1}
\end{eqnarray}
Substituting Eq.~(\ref{eq:Y^J_[0] y_+-}) into Eq.~(\ref{eq:L_sigma}), we can evaluate $L_\pm^{(1)}$ by iteration. Keeping up to the order
$r_\star^{-2}$, we obtain  
\begin{eqnarray}
    \frac{L^{(1)}_\pm }{ H_\pm } = 
    - \frac{ \alpha_\pm }{ 2N^3 } \Delta 
     \left[
       \Bigl( \frac{H_\pm}{a^2_\pm N} \Bigr) \Phi_\pm \pm  \sum_\sigma \Phi_\sigma  \right]  .
\label{eq:L^1 explicit}
\end{eqnarray}
Then Eq.~(\ref{eq:Y^J_[0lambda] y_+-}) is evaluated by using Eq.~(\ref{eq:Ninv}) as 
\begin{eqnarray}
  Y^{(1)}_{ \lambda -}  (r,y_-) 
  &\approx& 2N   \left[
   - \frac{ L_+^{(1)} }{H_+}
   + \frac{ L_-^{(1)} }{H_-}\left(1+\frac{a^2_+ H_-}{a^2_- H_+}\right)\right] 
   = \beta_-  O (r_\star^2 {\Delta  \Phi_-} ) 
   \bigl[  a_-^4 \alpha_+    +  \alpha_-  \bigr]
  \,,
\label{eq:Y^1_lam - alpha1a}\\
  Y^{(1)}_{ \lambda +}  (r,y_+) 
  &\approx& 2N  \left[
      \Bigl(\frac{\dot H_+}{3H_+^2} \Bigr)  \frac{ L_+^{(1)} }{H_+}
   +  \frac{ L_-^{(1)} }{H_-}\right]   
   = \beta_+  O (r_\star^2 {\Delta  \Phi_+} ) 
     \left[ 
      \Bigl(\frac{\dot H_+}{3H_+^2} \Bigr)^2 \alpha_{+}+\alpha_-\right] \,. 
\label{eq:Y^1_lam - alpha1b} 
\end{eqnarray}
On the negative tension brane, this new correction becomes important compared to that from the short-ranged part when the factor in the square brackets on the right hand side exceeds $O(a_-^2/m_S^2 \ell^2)$.
In particular, when $a_-^4 \alpha_+\gg 1$, Eq. (\ref{eq:m2limit}) applies and we find  $m^2\approx 4N^4/(3 \alpha_+ |\dot H_+|)\ll a_-^4 H^2_+/ ( \ell^{2}|\dot
H_+|)$. 
Because of the factor $a_-^4$, the mass of the stabilized radion becomes even   smaller. 
On the positive tension brane, the first term in the square brackets on the right hand side is suppressed only by the factor $\dot H_+/H_+^2$, which is small but is not hierarchically suppressed. 
As long as $\lambda_+$ takes the natural order of magnitude smaller than
$\ell$, $\alpha_+$ is at most $O(H_+^2/\dot H_+)$. 
Then, the correction to 
$Y_{ \lambda +}(r,y_+)$ stays less than 
$O(\beta_{+})O(r_\star^2 \Delta \Phi_+)$.
However, when $\lambda_+$ is much larger than $\ell$, the correction becomes larger by the factor of $\lambda_+/\ell$ than that in the strong coupling limit. 
Although these choices of parameters are not natural, the possibility of the enhanced correction without changing the order of radion mass might be interesting.

\subsection{Second order perturbation}

\subsubsection{Temporal component}

Let us discuss the contribution due to interaction terms to second order perturbations. 
In the strong coupling limit, the leading terms in second order perturbations are shown to be identical to that given by 4D Einstein gravity;
\begin{eqnarray}
  \bigl[ \Delta \bar A^{(2)} _{0\pm} \bigr]_{SC}
  =   8 \pi G(\rho_\pm^{(2)} +3P_\pm^{(2)}) 
    - 4 \Phi_\pm \Delta \Phi_\pm  
    + O\left( \frac{1}{a_-^2 a_{\pm}^2 r_\star^4} \right)
    \,. 
\end{eqnarray}
We will show that the corrections due to interaction terms are similarly suppressed as $O(1/a_{-}^2 a_{\pm}^2 r_\star^{4})$, where and hereafter we assume that $\lambda_\pm$ is not hierarchically enhanced and hence $\alpha_\pm$ is at most $O(1)$. 
In the following discussion we concentrate on the terms of $O(r_\star^{-4})$ with respect to the derivative expansion, neglecting the higher order terms than $O(r_\star^{-6})$. 
For simplicity, we adopt $\stac{(1)\,}{\hat \xi^r}=0$ as a choice of radial gauge in linear perturbations, keeping second order perturbations still in the isotropic gauge.

We quote the dependence of the first order perturbation variables on the warp factor from Eq.~(I-5.4): 
\begin{eqnarray}
A_{0\pm}^{(1)} \sim a^0, ~~~~
Y_{ \mathrm{pse} \pm}^{(1)} \,, ~ 
\varphi_{ \mathrm{pse}  \pm}^{(1)}  \sim \frac{a_{\mp}^2}{a^2}+1, ~~~~
\hat\xi^y_{\pm} \sim \frac{1}{a_{\pm}^2} \,. 
\label{eq:count a r 1}
\end{eqnarray}  
As for $u_\pm$ and $v_\pm$, we have Eqs.~(I-4.11) and (I-4.12) as 
\begin{equation}
  u_{\pm}  \sim \frac{a_{\mp}^2}{a^2}+1\,, ~~~~
  v_{\pm}  \sim \frac{a_{\mp}^2}{a^2}-1\,.
\label{eq:count uv}
\end{equation}

We list the perturbation variables of the first order that have a correction due to non-vanishing $\lambda_\pm$. 
As we have discussed, there is correction to $Y^{(1)}$, which is denoted by $Y^{(1)}_{ \lambda }$ in Eq.~(\ref{eq:Y=Y_0+Y_S}). 
Since $\varphi^{(1)}$ is related to $Y^{(1)}$ by Eq.~(I-2.13), 
$\varphi^{(1)}$ also has correction accordingly, which we denote by $\varphi^{(1)}_{ \lambda }$.
In the following discussion, the values of $\varphi^{(1)}_{ \lambda }$ evaluated at $y=y_\pm$ are necessary. They are obtained from Eq.~(\ref{eq:Jun of varphi}) as 
\begin{equation}
\varphi_{ \lambda }^{(1)}(y_\pm)
     = \mp\frac{3\lambda_\pm}{2\kappa a_\pm^2\dot\phi_{0\pm}}
         \Delta  Y^{(1)}_{ \mathrm{pse} }(y_\pm)
     + O\left( \frac{1}{r^4_\star} \right) . 
\label{eq:varphi_lambda}
\end{equation}
Although the source terms for second order perturbations are mostly written in terms of the variables in the Newton gauge, the expression $\bar A^{(1)}_{,y}$  in Gaussian normal coordinates is also necessary. 
From Eq.~(I-2.31), we can read the interaction term in 
$\bar A^{(1)}_{,y}$ as
\begin{equation}
\partial_y\bar A^{(1)}_{ \lambda }=\mp \frac{2\kappa \dot{\phi}_0}{3}
         \varphi_{ \lambda }^{(1)}
 \quad   ({\mathrm{at~}} y=y_\pm) , 
\end{equation}
where we used 
$\bar \varphi_{ \lambda }^{(1)}(y_\pm) = \varphi_{\lambda}^{(1)}(y_\pm)$, which follows from the fact that the gauge transformation  (I-2.26) of $\varphi^{(1)}$ is not altered by the effect of non-vanishing $\lambda_\pm$. 
Also for the variables $Y_{ \lambda }^{(1)}$ and
$\varphi^{(1)}_{ \lambda }$, we give the dependence on $a_-$ and $r_\star$.  
From Eqs.~(\ref{eq:Y^J_[0lambda] y_+-}) and (\ref{eq:varphi_lambda}), we obtain 
\begin{eqnarray}
    Y_{ \lambda -}^{(1)}(y_\pm),~
    \varphi_{ \lambda -}^{(1)}(y_\pm)  
    \sim \frac{1}{a_\pm^4 r_\star^2}. 
\end{eqnarray}
For later use, we quote the relations (I-B12) and (I-B13) in the notation of present paper as 
\begin{eqnarray}
  Y^{(1)}_{ \mathrm{pse} } + 2H_\pm \stac{(1)}{\hat\xi^y_\pm}  
     &=& \frac{2}{3} \sum _{\sigma=\pm }  \Phi_\sigma  
     \quad (\mathrm{at~} y=y_\pm)
\,, 
\nonumber
\\
    Y^{(1)}_{ \mathrm{pse} }   
    + \frac{2H_\pm}{\dot\phi_{0} } 
       \varphi^{(1)}_{ \mathrm{pse} }  
    &=& \frac{2}{3} \sum_{\sigma=\pm}  \Phi_\sigma   
       \quad (\mathrm{at~} y=y_\pm)
\,. 
\label{eq:comb Y varphi} 
\end{eqnarray}

We begin with the case that matter fields are confined on the negative tension brane. 
We discuss the contributions from each term in Eq.~(\ref{eq:bar A_0}) one by one. 
The second term on the right hand side of Eq.~(\ref{eq:bar A_0}) gives  
\begin{eqnarray}
 -\Delta \Bigl[  {\hat\xi_{-}^y} \bar{A} _{-,y} 
   + {\hat\xi_{-}^r} \bar{A} _{-,r}
   + \dot H ({\hat\xi_{-}^y})^2 
  \Bigr]
=  
   \bigl[ \cdots \bigr]_{SC}
 + \frac{2\kappa \dot\phi_{0-}}{3} 
   \Delta \bigl[\hat\xi^y_-  \varphi_{\lambda-}(y_-) \bigr]
 + O\left(\frac{1}{a_-^2 r_\star^4}\right) .
\label{eq:A2 part-gauge}
\end{eqnarray}
The contribution from the source terms $S_\xi$ and $S_\Sigma$ is evaluated by using the expressions given in Eqs.~(I-2.33) and (I-2.35) as
\begin{eqnarray}
  2N \sum_\sigma \sigma a^4_\sigma ( S_\xi^\sigma - S_\Sigma^\sigma)
=  
   -2N a^4_- \bigl[ S_\xi^- - S_\Sigma^- \bigr]_{SC}
  + O \left(\frac{ \Phi_-}{r_\star^2}Y_{ \lambda-}(y_-)  \right)
  + O\left(\frac{1}{a_-^2 r_\star ^4}\right) .
\end{eqnarray}
As for $(S_A - S_\psi)$, the underlined terms in Eqs.~(I-2.16) and
(I-2.19) give the leading correction as 
\begin{eqnarray}
2N\int_{y_+}^{y_-} a^2  (S_A - S_\psi)
 =   [\cdots]_{SC} 
  + O \left(\frac{\Phi_-}{r_\star^2}Y_{ \lambda -}(y_-)  \right)
  + O\left(\frac{1}{a_-^2 r_\star ^4}\right) , 
\end{eqnarray}
where we have used the fact that $\varphi_{\lambda-}(y_-)/\dot\phi_{0-}
= O(H^{-1}_- Y_{ \lambda-}(y_-))$. 
The leading order correction from $\mathbb{S}_\varphi$ comes from the last two terms in Eq.~(I-2.18). These terms are rewritten as in Eq.~(I-5.7), in which we did not assume the strong coupling limit. The underlined terms on the right hand side of 
Eq.~(I-5.7) give the leading order correction as  
\begin{eqnarray}
 2N \int^{y_{-}}_{y_{+}} dy ~ 
  a^2 v_{-} \Delta {\mathbb S}_{\varphi} 
  &=& 
 [\cdots ]_{SC}  
 - \frac{2\kappa}{3}\left[  \varphi_{ \mathrm{pse} -} \varphi_{ \lambda -} \right]_{y=y_-}
 - 2N \Delta \int dy 
  \left[ \frac{a^2 u_-}{2} \bigl(Y^2 - Y^2_{ \mathrm{pse} } \bigr)  \right] 
\nonumber
\\
 && 
   + O \left(\frac{\Phi_-}{r_\star^2} Y_{ \lambda -}(y_-)  \right)
   + O \left(\frac{1}{a_-^2 r_\star ^4}\right)\,.
\label{eq:A2 part-S_varp}
\end{eqnarray}
The last term cancels the contribution from $\mathbb{S}_Y$ that is given by  
\begin{eqnarray}
\Delta {\mathbb{S}}_Y = [\Delta {\mathbb{S}}_Y ]_{SC} 
   + \frac{\kappa}{3}a^2\dot\phi_0^2 \Delta 
    \bigl(Y^2 - Y^2_{ {\mathrm{pse}}} \bigr)  
  + O\left(\frac{1}{a_-^2 r_\star ^4}\right) \,,
\end{eqnarray}
where again the leading correction comes from the underlined terms in Eq.~(I-2.21). 
The second term in Eq.~(\ref{eq:A2 part-gauge}) and that in 
Eq.~(\ref{eq:A2 part-S_varp}), which potentially give enhanced correction to $\bar A^{(2)}_0$ of 
$O(\ell^2 \Phi_-^2/r_{\star}^2 a_-^6)$, cancel each other with the aid of Eq.~(I-3.27).  
The terms of $
O \left(r_\star^{-2}\Phi_- Y_{ \lambda -}(y_-)  \right)$ are smaller by the factor of $\Phi_-$ than the correction we have found for linear perturbations.

Now we consider the last term of Eq.~(\ref{eq:bar A_0}), i.e., the contribution  from $Y^{(2)}_{ \lambda }$.
To evaluate $Y^{(2)}_{ \lambda }$, we study $L_-$ given in Eq.~(\ref{eq:L_sigma}). 
According to the dependence (\ref{eq:count a r 1}), some terms have possibility to give an enhanced correction to $\Delta
Y_{ \lambda }^{(2)}$ of order $O(1/a_-^6 r^4_\star)$. 
Note that the terms of $O(1/a_-^4 r^4_\star)$ in $\Delta L_-$ give the terms of $O(1/a_-^6 r^4_\star)$ in $\Delta
Y_{ \lambda }^{(2)}$, while the contribution from $\Delta L_+$ does not change its order with respect to $a_-$. 
In the following discussion, we keep only the relevant terms that might give enhanced contributions of $O(1/a_-^6 r^4_\star)$ to $\Delta
Y_{ \lambda }^{(2)}$.

Keeping the terms of $O(1/a_-^6 r^2_\star)$, the source term $S^{-}_{jun}$ in Eq.~(\ref{eq:L_sigma}) becomes
\begin{eqnarray}
  \frac{ \lambda_{-}}{2\dot\phi_0} S^{-}_{jun} 
  - \frac{ \ddot\phi_0  \varphi^2}{2 \dot\phi_0^3} 
  + \frac{3 \varphi  \Delta  Y }
      {2\kappa a^2_- \dot\phi_0^3}
\approx 
  -\lambda_- 
\left[
   \frac{(\hat\xi^y_{-,r})^2}{2a_-^2}
+  \frac{3}{2\kappa a_-^2 \dot\phi_{0-}^2 }
\Bigl(  {\mathbb S}_Y  + a_-^2 \dot H  Y^2\Bigr)
\right]_{y=y_-} , 
\label{eq:Sjun terms}
\end{eqnarray}
where we used the expression given in Eq.~(I-B18). 
Here the term of ${\mathbb S}_Y$ is evaluated by using Eq.~(I-2.21). Keeping the terms of $O(1/a_-^4 r_\star^2)$, we obtain
\begin{eqnarray}
{\mathbb S}_Y  + a_-^2 \dot H  Y^2
&\approx&
 - \int \biggl[
  3 Y_{,r} \Delta Y  
 + \frac{2H\varphi}{\dot \phi_0} (\Delta Y)_{,r} 
 + \frac{2 \kappa}{3} 
    \bigl(
       \varphi_{,r} \Delta \varphi 
       - \varphi (\Delta \varphi)_{,r} 
    \bigr)
  +4 a^2 B_{,y} 
  \Bigl( 2 H Y_{,r}+ \dot H \frac{\varphi_{,r}}{\dot\phi_0}  \Bigr) 
      \biggr]  dr \,,
\end{eqnarray}
where we have used Eq.~(\ref{eq:comb Y varphi}).
The right hand side of Eq.~(\ref{eq:L_sigma})  also contains $Y^{(2)}$. 
To obtain the lowest order correction, this $Y^{(2)}$ can be replaced with $Y_{ \mathrm{pse} }^{(2)}$. 
We can neglect the contribution from 
$Y_{ \mathrm{pse} }^{(2)}(r,y_{+})$ in Eq.~(\ref{eq:Y^J_[0lambda] y_+-}), which is not enhanced with respect to the factor of $a_-$. 
The relevant terms in $\Delta Y_{ \mathrm{pse} }^{(2)}(r, y_{-})$ are $O(1/a_-^4 r_\star^2)$, which are 
\begin{eqnarray}
 \Delta Y^{(2)}_{ \mathrm{pse}} (r, y_{-})
 \approx 
    \frac{\kappa}{3} \Delta ( \varphi^2_- )
 -  \Bigl( S_\psi + 2H a_{-}^2 S_{\xi}^{-} \Bigr), 
\label{eq:Y^2 terms}
\end{eqnarray}
where the first term comes from the integration of ${\mathbb
S}_\varphi$.  
Using Eq.~(I-2.16), $S_\psi$ in the second term can be explicitly written down. 
The relevant contribution comes from only the underlined terms in Eq.~(I-2.16). 
With the aid of Eqs.~(I-2.10),  (I-2.13), (I-B14) and the fact that $(a^4 B_{,y})_{,y}=O(a^2/r_\star^2)$, we obtain 
\begin{eqnarray}
 S_\psi 
&\approx&
  \int dr 
  \biggl\{
    4a^2 B_{,y} \Bigl(HY_{,r} +\dot H  \frac{\varphi_{,r}}{\dot \phi_0}\Bigr)
  + \frac{1}{ r^{8/3}} \partial_r
 \Bigl[ r^{8/3} 
  \Bigl( \frac{3}{2} Y_{,r}^2 + \kappa \varphi_{,r}^2 
  \Bigr)
  \Bigr] 
 \biggr\}  \,. 
\end{eqnarray}
As for $S_\xi^-$, we can read from Eq.~(I-2.35) as 
\begin{eqnarray}
 a_{-}^2S_\xi^-
&\approx & 
 2  \int \Bigl[ a_{-}^2  B_{,y} Y_{,r}  
 + \hat \xi^y_- (\Delta Y)_{,r} \Bigr] dr 
 -   \hat\xi_{-}^y \Delta Y   
 - H  \bigl( {\hat\xi_{-,r}^y} \bigr)^2 ,
\end{eqnarray}
where we have again used Eqs.~(\ref{eq:comb Y varphi}).
Substituting all the results into Eq.~(\ref{eq:L_sigma}), we obtain 
\begin{eqnarray} 
\frac{L_-^{(2)}}{H_-} 
\approx 
 \frac{3\alpha_- }{ 4 N^3 }
 \int dr \biggl[
   \frac{ \kappa}{3} ( \varphi_{,r}^2 -  (\dot\phi_0 \hat\xi^y_{,r})^2)_{,r}
 + \frac{2H (\Delta Y)_{,r}}{\dot\phi_0} (\varphi - \dot\phi_0 \hat\xi^y) 
 + \frac{2}{r} \Bigl( Y_{,r}^2 -  (2H \hat \xi^y_{,r})^2 \Bigr)
 + 2H \hat\xi^y_{,r} \Delta (Y+2H\hat\xi^y)
 \biggr] .
\end{eqnarray}
Further application of Eqs.~(\ref{eq:varphi_lambda}) and (\ref{eq:comb Y varphi}) reduces the order with respect to either $a_-^{-1}$ or $r_\star$. 
Therefore we find 
$L_-^{(2)}= \alpha_- O(\ell^2 \Phi_-^2/r_{\star}^2 a_-^2)+O(1/r_{\star}^4)$. 
Although we have not discussed the contributions associated with the factor $\alpha_+$ in detail, it is manifest that they do not have any enhancement with respect to the hierarchy factor of $1/a_-$. 
Therefore the interaction terms in second order perturbations become
\begin{eqnarray}
    \Delta  \bar A^{(2)}_{0-}(r,y_-) 
 = \left[\Delta \bar A^{(2)}_{0-}(r,y_-)\right]_{SC}
 +  O\left( \frac{\beta_- \Phi_-^2}{ r_\star^{ 2}} \right)
    \left[O(a^4_-  \alpha_+) + O(\alpha_-)\right]\,,
\label{eq:barA2}
\end{eqnarray}
which are suppressed compared to the correction at the linear order by the factor of $\Phi_-$.

Next we consider the case that the matter distribution is concentrated on the positive tension brane. 
In this case, the first order quantities listed in
Eqs.~(\ref{eq:count a r 1}) and (\ref{eq:count uv}) do not suffer enhancement with respect to the factor of $1/a_-$. 
The factor $1/a_-$ arises only through $Y_{ \lambda +}(y_-)$ and $\varphi_{ \lambda +}(y_-)$, both of which are $O(1/a_-^2)$. 
Note also that the terms with $\lambda_+$ are always associated with $u_+(y_+)=O(\dot H_+/H_+^2)$. 
From these observations, we find 
\begin{eqnarray}
    \Delta  \bar A^{(2)}_{0+}(r,y_+) 
 = \left[\Delta \bar A^{(2)}_{0+}(r,y_+)\right]_{SC}
 + O\left( \frac{\beta_+ \Phi_+^2}{ r_\star^{ 2}} \right)
     \left[  \Bigl(\frac{\dot H_+}{ H_+^2} \Bigr)^2  O( \alpha_+ )
     +  O \Bigl( \frac{ \alpha_- }{a^2_-} \Bigr)\right]    \,. 
\label{eq:A2+}
\end{eqnarray}
The amplitude of this second order correction is not simply suppressed by the factor of $\Phi_+$ compared to the correction at the linear order. 
There is a difference with respect to the power of $a_-$ in the term with $\alpha_-$. 
In paper I, we have met a similar phenomenon in the analysis of correction due to Kaluza-Klein modes at the second order. As was discussed there, this phenomenon is a natural consequence of our gradient expansion approximation.

\subsubsection{Spatial component}
 
Finally, we comment on the spatial component. 
The last two terms on the right hand side of Eq.~(\ref{eq:bar B^2}) are evaluated in the same manner as was done for $\Delta \bar A_{0\pm}^{(2)}$. 
The source terms except for $S_B$ have been already computed in evaluating $\Delta \bar A_{0\pm}^{(2)}$.  
The contribution from $S_B$ is given by 
\begin{eqnarray}
\Delta S_B\approx \bigl[ \Delta S_B \bigr]_{SC}
- \kappa \dot\phi_0 \Delta \bigl[
  \hat \xi_\pm^y (\varphi - \dot \phi_0 \hat  \xi^y_\pm)
 \bigr]\,, 
\end{eqnarray}
which is similar to that given in Eq.~(\ref{eq:A2 part-gauge}). 
Combining all, we obtain
\begin{eqnarray}
  \bar B 
=
  \bigl[ \bar B \bigr]_{SC} 
+ O \left(  \bar A^{(2)}_{0 \pm} -[   \bar A^{(2)}_{0 \pm}]_{SC} 
    \right) \,,
\end{eqnarray}
and therefore the correction is the same order as that of the temporal component.

\section{Summary}
\label{sec:summary}

We have discussed metric perturbations in the Randall-Sundrum two branes model with radius stabilization.  As a mechanism for radius stabilization, we have assumed a scalar field with a potential in the bulk and that on each brane.  In our previous work (paper I), we took the strong coupling limit, in which the brane potential is extremely narrow so that the values of the scalar field on the branes cannot fluctuate. 
In this paper, we extended the previous analysis relaxing the limitation of taking the strong coupling limit.

In the strong coupling limit, it is known that the mass squared of the stabilized radion tends to be hierarchically small as $m_S^2=O((a_-/a_+)^2 \ell^{-2})$, where $\ell$ is the bulk curvature scale and $(a_-/a_+)$ is the ratio of the values of the warp factor on the respective branes. 
First we examined the shift of this mass eigenvalue when we relax the limitation of taking the strong coupling limit. 
We have shown that a tachyonic mode appears when either of the coupling constants $\lambda_\pm$ defined in Eq. (\ref{eq:lambdadef}) is negative. 
Hence, the models with such parameters are unstable. 
When both $\lambda_+$ and $\lambda_-$ are positive, we derived formulas for the mass squared of the stabilized radion in Eq.~(\ref{eq:m2limit}) for  
$\lambda_+ \gg m_S^{-2}\ell^{-1}$ and in Eq.~(\ref{eq:m2small}) for $\lambda_+ \ll m_S^{-2}\ell^{-1}$.  
The mass squared of the stabilized radion is affected by $\lambda_-$ when $\lambda_-\gtrsim \ell$ while 
$\lambda_+$ only when $\lambda_+ \agt m_S^{-2}\ell^{-1}$. 
These are rather exceptional cases since the order of 
$\lambda_\pm$ is typically less than or equal to $O(\ell)$.

Next, we have examined the effects on metric perturbations induced on the branes by matter fields up to second order, assuming that the matter distribution confined on the branes is static and spherically symmetric. 
For simplicity, we assumed that matter fields reside on either of the two branes. 
The results for the case that the matter fields are on the negative tension brane are summarized by Eqs.~(\ref{eq:Y^1_lam - alpha1a}) and (\ref{eq:barA2}), 
where $\alpha_\pm$ and $\beta_\pm$ are defined in 
Eq.~(\ref{eq:def. alpha}) and Eq.~(\ref{eq:suppres factor 1}), respectively. 
The correction due to the finite coupling becomes important compared to that already existing in the strong coupling limit only when the mass of the stabilized radion is significantly reduced by the effect of nonvanishing $\lambda_\pm$.
The results for the case that the matter fields reside on the positive tension brane are summarized by Eqs.~(\ref{eq:Y^1_lam - alpha1b}) and
(\ref{eq:A2+}).  
It is possible that the correction due to the finite coupling becomes important when $\lambda_+$ becomes much larger than $\ell$ without changing the mass of the stabilized radion. The corrections are enhanced by the factor of $\lambda_+/\ell$ compared to those present in the strong coupling limit. 
The result for second order perturbations (\ref{eq:A2+}) seems to show that the correction associated with $\lambda_-$ are enhanced by a factor of $a_+^2/a_-^2$. 
However, this is an artifact due to the limitation of the present approximation using gradient expansion method.

In conclusion, under the assumption that the matter distribution is static and spherically symmetric, we have confirmed that 4D Einstein gravity is approximately recovered up to second order perturbations relaxing the limitation of taking the strong coupling limit. 
The condition that the corrections due to $\alpha_\pm$, $\beta_\pm$, and radion mass are sufficiently suppressed gives a consistency check for any stabilization model to use the scalar field of Eq. (\ref{eq:Lagrangian}) \cite{Gibbons:2001tf}.

\begin{acknowledgments}
H.K. would like to thank Takashi Nakamura and Hideo Kodama for their continuous encouragement. To complete this work, the discussion during and after the YITP workshop YIYP-W-01-15 on ``Braneworld - Dynamics of spacetime boundary'' was useful. 
H.K. is supported by the Japan Society for the Promotion of Science under the Predoctoral Research Program. 
This work is partly supported by the Monbukagakusho Grant-in-Aid No.~1270154  and the Yamada Foundation.

\end{acknowledgments} 

\appendix

  

\end{document}